\documentclass[fleqn,11pt,twoside]{article}

\usepackage{amsthm,amsthm,amssymb, color, xcolor,epsfig, graphics, subfigure}

\usepackage{amsmath, graphicx, latexsym, lscape}

\makeatletter
\newcommand{\copyrightnote}[2]{{\renewcommand{\thefootnote}{}
 \footnotetext{\small\it
\begin{flushleft}
 \copyright \ #1   #2
\end{flushleft}}}}

\newcommand{\Name}[1]{\begin{flushleft}
                       \LARGE \bf #1
                       \end{flushleft}\vspace{-3mm}}

\newcommand{\Author}[1]{\begin{flushleft}
                       \it #1 \end{flushleft}}

\newcommand{\Address}[1]{\begin{flushleft}
                       \it #1 \end{flushleft}}

\newcommand{\Date}[1]{\begin{flushleft}
                      \small  \it #1 \end{flushleft}}

\newcommand{\evenhead}{Author \ name}
\newcommand{\oddhead}{Article \ name}

\renewcommand{\@evenhead}{
\hspace*{-3pt}\raisebox{-15pt}[\headheight][0pt]{\vbox{\hbox to \textwidth
{\thepage \hfil \evenhead}\vskip4pt \hrule}}}
\renewcommand{\@oddhead}{
\hspace*{-3pt}\raisebox{-15pt}[\headheight][0pt]{\vbox{\hbox to \textwidth
{\oddhead \hfil \thepage}\vskip4pt\hrule}}}
\renewcommand{\@evenfoot}{}
\renewcommand{\@oddfoot}{}

\long\def\@makecaption#1#2{%
  \vskip\abovecaptionskip
  \sbox\@tempboxa{\small \textbf{#1.}\ \ #2}%
  \ifdim \wd\@tempboxa >\hsize
    {\small \textbf{#1.}\ \ #2}\par
  \else
    \global \@minipagefalse
    \hb@xt@\hsize{\hfil\box\@tempboxa\hfil}%
  \fi
  \vskip\belowcaptionskip}

\newcommand{\JNMPnumberwithin}[3][\arabic]{%
  \@ifundefined{c@#2}{\@nocounterr{#2}}{%
    \@ifundefined{c@#3}{\@nocnterr{#3}}{%
      \@addtoreset{#2}{#3}%
      \@xp\xdef\csname the#2\endcsname{%
        \@xp\@nx\csname the#3\endcsname .\@nx#1{#2}}}}%
}

\newcommand{\resetfootnoterule} {
  \renewcommand\footnoterule{%
  \kern-3\p@
  \hrule\@width.4\columnwidth
  \kern2.6\p@}
}

\renewcommand{\footnoterule}{}

\makeatother

\theoremstyle{definition}
\newtheorem*{definition}{Definition}
\newtheorem*{example}{Example} 

\usepackage{amsmath,amsfonts,amssymb,amsthm,amsxtra,epsfig,epstopdf,url,array}

\usepackage{graphicx}
\usepackage{color}
\usepackage{latexsym}
\usepackage{bm}

\def\dh#1{\mathop {#1}\limits_{h}}

\def\dh#1{ \mathop{#1}\limits_ h}

\def\dh#1{\mathop {#1}\limits_h}

\def\dvhb#1{\dh#1_{\bar 1}}
\def\dvhb2#1{\dh#1_{\bar 2}}

\def\dh#1{\mathop {#1}\limits_{h}}

\def\dvhb#1{\dh#1_{\bar x}}

\def\dh#1{\mathop {{#1}}\limits_{h}}

\def\dh#1{ \mathop{#1}\limits_ h}

\newcommand{\DD}{D}   
\newcommand{\Dt}{D}    

\newcommand{\deltauL}{{\delta L \over \delta  u}}

\newtheorem{theorem}{Theorem}[section]
\newtheorem{lemma}[theorem]{Lemma}
\newtheorem{remark}[theorem]{Remark}
\newtheorem{corollary}[theorem]{Corollary}
\newtheorem{proposition}[theorem]{Proposition}

\newtheorem{ass}{Assumption}[section]

\numberwithin{equation}{section}

\begin{document}

\renewcommand{\evenhead}{ {\LARGE\textcolor{blue!10!black!40!green}{{\sf \ \ \ ]ocnmp[}}}\strut\hfill
V Dorodnitsyn, R Kozlov and S Meleshko
}
\renewcommand{\oddhead}{ {\LARGE\textcolor{blue!10!black!40!green}{{\sf ]ocnmp[}}}\ \ \ \ \
Symmetries and Integration of ODEs with Retarded Argument
}

\thispagestyle{empty}
\newcommand{\FistPageHead}[3]{
\begin{flushleft}
\raisebox{8mm}[0pt][0pt]
{\footnotesize \sf
\parbox{150mm}{{\textcolor{blue!10!black!40!green}{{\bf Open Communications in Nonlinear Mathematical Physics}}}
\ \ {Special Issue: Bluman}, 2025\\[0.1cm]
\strut\hfill
ocnmp:16819,
pp #2\hfill {\sc #3}}}\vspace{-13mm}
\end{flushleft}}

\FistPageHead{1}{\pageref{firstpage}--\pageref{lastpage}}{ \ \ }

\strut\hfill

\strut\hfill

\copyrightnote{The authors. Distributed under a Creative Commons Attribution 4.0 International License}

\begin{center}

{\bf {\large A Special OCNMP Issue in Honour of George W Bluman}}
\end{center}

\smallskip

\Name{Symmetry Approach to Integration of   \\
Ordinary Differential Equations\\
with Retarded Argument}

\Author{Vladimir  Dorodnitsyn$^a$\footnote{Corresponding author},
Roman Kozlov$^b$, Sergey  Meleshko$^c$}

\Address{$^a$Keldysh Institute of Applied Mathematics, Russian Academy of Science, \\
Miusskaya Pl.~4, Moscow, 125047, Russia; \\
{e-mail:  Dorodnitsyn@Keldysh.ru, dorod2007@gmail.com} \\
$^b$ Department of Business and Management Science, Norwegian School
of Economics,
Helleveien 30, 5045, Bergen, Norway;  \\
{e-mail: Roman.Kozlov@nhh.no}  \\
$^c$ School of Mathematics, Institute of Science,
Suranaree University of Technology, 30000, Thailand 
\\
{e-mail: sergey@math.sut.ac.th} 
}

\Date{Received October 30, 2025; Accepted November 2, 2025}

\setcounter{equation}{0}

\smallskip

\noindent
{\bf Citation format for this Article:}\newline
V Dorodnitsyn, R Kozlov and S Meleshko,
Symmetry Approach to Integration of Ordinary Differential Equations with Retarded Argument,
{\it Open Commun. Nonlinear Math. Phys.}, Special Issue:\,Bluman, ocnmp:16819, \pageref{firstpage}--\pageref{lastpage}, 2025.

\strut\hfill

\noindent
{\bf The permanent Digital Object Identifier (DOI) for this Article:}\newline
{\it 10.46298/ocnmp.16819}
\strut\hfill

\begin{abstract}

\noindent
We review studies on the application of Lie group methods to delay ordinary differential equations (DODEs). For first- and second-order DODEs with a single delay parameter that depends on independent and dependent variables, the group classifications are performed. Classes of invariant DODEs for each Lie subgroup are written out. The symmetries allow us to construct invariant solutions to such equations. The application of variational methods to functionals with one delay yields DODEs with two delays. The Lagrangian and Hamiltonian approaches are reviewed. The delay analog of the Legendre transformation, which relates the Lagrangian and Hamiltonian approaches, is also analysed. Noether-type operator identities relate the invariance of delay functionals with the appropriate variational equations and their conserved quantities. These identities are used to formulate Noether-type theorems that give first integrals of second-order DODEs with symmetries. Finally, several open problems are formulated in the Conclusion.

\end{abstract}

\label{firstpage}


\section{Introduction}

Since its inception in the classical works of Sophus Lie~\cite{bk:Lie1888, bk:Lie1924}, Lie group analysis has proven to be a powerful and effective tool for studying both ordinary and partial differential equations. The Lie group symmetries of a differential equation map solutions to other solutions, enabling the generation of new solutions from known ones and facilitating the classification of equations into equivalence classes. Moreover, these symmetries can be used to derive exact analytical solutions that remain invariant under specific subgroups of the symmetry group, commonly referred to as group-invariant solutions.

The application of Lie group theory to differential equations has been extensively developed in numerous books and research articles~\cite{bk:Ovsiannikov1978, bk:Ibragimov1983, bk:Olver1986, bk:Gaeta1994, bk:HandbookLie, bk:BlumanAnco2002}. Following the seminal contribution of E. Noether~\cite{Noether1918}, symmetry groups have also served as a fundamental tool for deriving first integrals and conservation laws in systems with Lagrangian or Hamiltonian formulations. Furthermore, the connection between symmetry groups and conservation laws for differential equations that lack a variational structure -- hence without an associated Lagrangian or Hamiltonian -- was established in~\cite{bk:AncoBluman1997, bk:BlumanAnco2002}.

The application of Lie group transformations has been successfully extended beyond continuous systems to encompass finite-difference, discrete, and differential-difference equations \cite{Dorodnitsyn1991, LeviWinternitz1991, QuispelCapelSahadevan, Dorodnitsyn1993,  Dorodnitsyn1993a, DorodnitsynKozlovWinternitz2000, LeviWinternitz2005, bk:Dorodnitsyn2011, bk:Levi_book_2011, bk:Hydon2014, DKKW2015,  bk:Levi2023} as well as integro-differential equations~\cite{bk:GrigorievIbragimovKovalevMeleshko2010}. In addition, exact solutions of delay partial differential equations (PDEs) have been investigated in~\cite{Polyanin_Zhurov_2014b, Polyanin_Zhurov_2023, Polyanin2023}.


The present article reviews the extension of Lie group analysis methods to delay ordinary differential equations (DODEs). In earlier studies, DODEs with one delay were considered. A Lie group classification of first-order delay ordinary differential equations was presented in~\cite{DKMW2018a}, while linear first-order delay ordinary differential equations were examined in~\cite{DKMW2018b}. Furthermore, a Lie group classification of second-order delay ordinary differential equations was developed in~\cite{DKMW2021}. Results concerning the possible dimensionality of the admitted Lie algebras were generalized to DODEs of arbitrary order in~\cite{DKMW2019}.

A Lagrangian approach to DODEs and a Noether-type theorem for invariant variational problems were recently presented in~\cite{DKM2023}. Corresponding results for a Hamiltonian approach to DODEs were suggested in~\cite{DKM2024arXiv, DKM2025}. We remark that Lagrangian and Hamiltonian formalisms for difference equations were discussed in \cite{bk:DorodnitsynKozlov2011} (see also  \cite{bk:DorodnitsynKozlov2009, bk:DorodnitsynKozlov2010}).

A variational approach to delay differential equations was introduced long ago in~\cite{Elsgolts2, Elsgolts3, bk:Elsgolts1955, Huges1968, Sabbagh1969}; however, the symmetries of the corresponding variational equations were not investigated until recently. For simplicity, we restrict our attention to the scalar case, considering functions involving a single delay, which correspond to two time points: the current time point  $t$ and the delayed time point $t -\tau $, where $\tau $ denotes the delay parameter. Such functionals give rise to variational delay ordinary differential equations of second order (in terms of the order of differentiation) that involve two delays: for each time point $t$, two delayed arguments $ t - \tau $ and  $ t - 2 \tau $ appear. Consequently, the initial data must be specified over an interval of length $ 2 \tau$.

The Noether theorems for variational delay ordinary differential equations represent generalizations of Noether’s classical theorems for both variational ordinary differential equations and variational ordinary difference equations. The DODE analogues of Noether’s theorems provide a systematic method for deriving first integrals directly from the Lagrangian and Hamiltonian functions. When a sufficient number of first integrals is obtained, they can be employed to construct explicit solutions of the corresponding DODEs.

The structure of the paper is as follows. In the next section, we review results concerning Lie group symmetries of DODEs with one delay. Section \ref{section_DODE} introduces second-order delay ordinary differential equations involving two delays. Section~\ref{section_Lie_point} presents the application of Lie point symmetry generators to DODEs. Sections  \ref{Lagrangian_framework} and  \ref{Hamiltonian_framework} provide Lagrangian and Hamiltonian approaches for DODEs. They review the variational equations in the continuous case and their generalizations for DODEs, invariance of delay functions, and Noether-type theorems. The final section provides a concise summary of the main results and lists several directions for future research.

\section{Delay ordinary differential equations with one delay}

\subsection{First-order DODEs}

The adaptation of  Lie group and Lie algebra theory to the study of delay ordinary differential equations started in~\cite{DKMW2018a, DKMW2018b}. In these articles, we restricted ourselves to the case of first-order DODEs, supplemented by a general delay equation. Thus, we consider a {\it delay ordinary differential system} (DODS) of the from
\begin{subequations}
\label{first_order}
\begin{gather}
\label{first_order_a}
\dot{u} = f ( t, u, u_-  ),
\qquad
t \in I,
\qquad
{ \partial f  \over  \partial u_- }  {\not\equiv}  0,
\\
\label{first_order_b}
t_- = g ( t, u, u_-   ),
\qquad
t_- <  t,
\qquad
g  ( t, u, u_-  ) {\not  \equiv}   \mbox{const}.
\end{gather}
\end{subequations}
Here, $I$ is a finite or semi-finite interval and $f$ and $g$ are arbitrary smooth functions.

Sometimes it is convenient to rewrite~(\ref{first_order_b}) in the equivalent form
\begin{equation*}
\Delta t  = t - t_- = \tilde{g} ( t, u, u_- ),
\qquad
\tilde{g} ( t, u, u_- ) =   t -  g ( t, u, u_-  ).
\end{equation*}
In most of the existing literature, the delay parameter $ \Delta t $ is considered to be constant
\begin{equation}    \label{consntantdelay}
 \Delta t  = \tau  > 0,
 \qquad
 \tau  = \mbox{const}.
\end{equation}
An alternative is to impose a specific form of the function~(\ref{first_order_b}) to include some physical features of the delay $\Delta t$.

We are interested in group transformations that leave the Eqs.~(\ref{first_order}) invariant (Generators of Lie group transformations are described in section \ref{section_Lie_point}, where DODEs with two delays are considered). That means that the transformation will transform solutions of the DODS into other solutions. They leave the set of all solutions invariant. Let us stress that we need to consider two equations~(\ref{first_order}) together. It makes our approach similar to one of the approaches for considering symmetries of discrete equations (see, for example,~\cite{bk:Dorodnitsyn2011}), where invariance is required for both the discrete equation and the equation specifying the lattice on which it is considered. Here, we have the DODE~(\ref{first_order_a})  instead of a discrete equation and the delay relation~(\ref{first_order_b}) instead of a lattice equation.

\medskip

The main results were the following:

\begin{enumerate}

\item

We classified DODSs of the form~(\ref{first_order})  into conjugacy classes under arbitrary Lie point transformations and found that their Lie point symmetry groups can have dimensions $ n=0, 1, 2, 3$ or $n=\infty$.

\item

The symmetry algebras for genuinely nonlinear DODEs have dimensions $n \leq 3$.

\item

If $n=\infty$, the DODE is linear and we have a solution-independent delay equation given by $g=g(t)$, or it can be transformed into this linear form by a point transformation.

\item

If the symmetry algebra of a DODS contains a 2-dimensional {sub}algebra realized by linearly connected vector fields,  then this DODS  is linearizable (or already linear with $g=g(t)$).

\item

We presented a method for obtaining particular solutions for DODSs with symmetries. Exact analytical solutions of invariant DODSs can be obtained using symmetry reduction.

\end{enumerate}

\begin{example}

Let us consider a DODE  with a constant delay parameter
\begin{equation}        \label{Illustrating_DODE}
 { \displaystyle   \dot{u}  =    {  u - u_-   \over  t - t_-   }   +  C_1   e^t  },
\qquad
t  -t _-   = C_2,
\qquad
C_2   >   0.
\end{equation}

These equations admit symmetries
\begin{equation}        \label{Illustrating_symmetries}
 { \displaystyle
X_1 =     {\partial  \over \partial u },
\quad
X_2 = t  {\partial  \over \partial u },
\quad
X_3 =     {\partial  \over \partial t }    +  u   {\partial  \over \partial u }      }.
\end{equation}
One can easily see that the corresponding Lie group transformations
\begin{equation*}
\begin{array}{lll}
X_1 :
&
\bar{t} = e^{ \varepsilon X_1 } (t) = t,
&
\bar{u} = e^{ \varepsilon X_1 } (u) = u  +  \varepsilon,  \\
X_2 :
&
\bar{t} = e^{ \varepsilon X_2 } (t) = t,
&
\bar{u} = e^{ \varepsilon X_2 } (u) = u  +  \varepsilon t,  \\
X_3 :
&
\bar{t} = e^{ \varepsilon X_3 } (t) = t  +  \varepsilon,
&
\bar{u} = e^{ \varepsilon X_3 } (u) =   e^{ \varepsilon} u   \\
\end{array}
\end{equation*}
do not change the form of the DODE, nor the delay relation.

We can find a particular solution of~(\ref{Illustrating_DODE}), which is an invariant solution for the symmetry $X_3$. It has the form
\begin{equation}        \label{Illustrating_form}
u =   A    e^t,
\qquad
t_-  = t - B,
\end{equation}
where  $A$ and $B$  are constants. This solution form is invariant with respect to the group transformation generated by the symmetry $X_3$. Substituting the form~(\ref{Illustrating_form}) into the DODS~(\ref{Illustrating_DODE}), we obtain the conditions on the constants
\begin{equation*}
A = {  { C_1 C_2 } \over  {C_2 - 1 +  e ^{ - C_2 }  }  },
\qquad  B  =  C_2.
\end{equation*}

\end{example}

Other examples of invariant solutions can be found in~\cite{DKMW2018a, DKMW2018b,  DKMW2019}.

\subsection{Second-order DODEs}

Second-order delay ordinary differential equations with one delay were examined in~\cite{DKMW2021}. These DODSs have the form
\begin{subequations}
\label{DODS}
\begin{gather}
 \label{DODE}
\ddot{u} = f ( t, u, u_-, \dot{u}, \dot{u} _-   ),
\qquad
t \in I,
\qquad
\left( { \partial f  \over  \partial u_- } \right) ^2    +   \left( { \partial f  \over  \partial \dot{u}_- } \right) ^2
{\not\equiv}  0,
\\
 \label{delay}
t_- = g ( t, u, u_-, \dot{u}, \dot{u} _-   ),
\qquad
t_- <  t,
\qquad
g  ( t, u, u_-, \dot{u}, \dot{u} _-  ) {\not  \equiv}   \mbox{const}.
\end{gather}
\end{subequations}
Here $f$ and $g$ are arbitrary smooth functions.

\medskip

The following results were obtained:

\begin{enumerate}

\item

A second-order genuinely nonlinear DODS of the type (\ref{DODS}) can have a Lie point symmetry algebra of dimension $ 0 \leq n  \leq 6$ or $ n = \infty$.

\item

Genuinely nonlinear DODEs have symmetry algebras with dimensions $n \leq 6$.

\item

If we have  $n = \infty$, then the DODS is linearisable by an invertible transformation. The delays of linear DODEs with $n = \infty$ are solution-independent, i.e., given by $ g = g(t)$.

\item

To show that a DODS (\ref{DODS}) is linearisable (and its delay equation can be brought to the form $ g = g(t)$), it is sufficient to show that it admits a four-dimensional symmetry algebra realized by pairwise linearly connected vector fields.

\item

Particular solutions of invariant DODSs can be obtained using symmetry reduction.

\end{enumerate}

\subsection{Higher  order DODEs}

In~\cite{DKMW2019}, some results obtained for the first- and second-order DODSs in~\cite{DKMW2018a, DKMW2018b,  DKMW2021} were generalized for DODEs of arbitrary order with one delay. Such systems of order $N$  have the form
\begin{subequations}  \label{N_DODS}
\begin{gather}   \label{N_DODE}
{u}^{(N)}  = f ( t, u, u_-,  \dot{u}, \dot{u}_-,...,  u^{(N-1)}, u_- ^{(N-1)}),
\qquad
t \in I,
\qquad
\sum _{k = 0 } ^{N-1}
\left( { \partial f  \over  \partial  u^{(k)}  _-  }  \right) ^2   {\not\equiv}  0,
\\
  \label{N_delay}
t_- = g  ( t, u, u_-,  \dot{u}, \dot{u}_-,...,  u^{(N-1)}, u_- ^{(N-1)}),
\qquad
t_- <  t,
\qquad
g  {\not  \equiv}   \mbox{const}.
\end{gather}
\end{subequations}

The dimension of a symmetry algebra admitted by a DODS of order $N$ can satisfy $0 \leq n \leq 2N + 2$ or $n = \infty$. Genuinely nonlinear DODS can admit symmetry algebras of dimension $0 \leq n \leq  2N + 2$. If $n = \infty$, the DODS consists of a linear DODE and a solution-independent delay relation, or it can be transformed into such a form by an invertible transformation.

For invariant DODSs, we obtained several theoretical results. Namely, if the symmetry algebra has $2N$ pairwise linearly connected symmetries, it provides a delay differential system which can be transformed into a linear DODE supplemented by a solution-independent delay relation. Such linear DODSs admit infinite-dimensional symmetry groups since they allow linear superposition of solutions.


\section{Delay ordinary differential equations with two delays}

\label{section_DODE}

In sections  \ref{Lagrangian_framework} and  \ref{Hamiltonian_framework}, we consider variations of delay functionals with one delay. Such variations provide delay ordinary differential equations with two delays. In the Lagrangian framework, we obtain second-order DODEs. For this reason, we describe the second-order DODEs with two delays.

\subsection{Second-order DODEs}

Second-order DODEs with two delays have the form
\begin{subequations}
 \label{DODE0}
\begin{gather}
\label{DODE_EQ}
\ddot{u} ^+ =
F (
t ^+,  t, t^-,
u^+, u, u^{-},
\dot{u}^+, \dot{u},\dot{u}^{-},
\ddot{u},\ddot{u}^-
),
 \qquad
t \in I,
\\
   \label{delays}
t  ^+  -   t   =  \tau,
\qquad
t  -   t ^-  =  \tau,
\qquad
\tau = \mbox{const}.
\end{gather}
\end{subequations}
where $I \subset  \mathbb{R} $ is some finite or semi{finite} interval.
The independent variable $t$ varies continuously over the entire region, where Eq.~(\ref{DODE0}) is defined\footnote{In the literature on  DODEs with two delays, it is standard to consider three points $t$,   $t - \tau$, and $t - 2 \tau$. We prefer to use three points  $t^+  = t  + \tau$, $t$, and $t^- = t   - \tau$, as this choice is more suitable for variational delay equations.}. We emphasize that the delay parameter takes the same value $\tau$ at all points. It is convenient to use the right and left shift operators, defined for a function
$ f = f ( t, u) $ as
\begin{equation}  \label{shifts}
f^+ = S_+ (f)   = f(t+\tau, u (t  + \tau)  ),
\qquad
f^- = S_{-} (f)= f(t-\tau, u (t  - \tau) ).
\end{equation}
For example,
\begin{equation*}
t^+   = S_+ (t) = t + \tau,
\qquad
  t^{-}= S_- (t) =  t -  \tau
\end{equation*}
and
\begin{equation*}
u^+ =   S _+ ( u ) =   u(t + \tau),
\qquad
u^{-}  =   S_- (u) =   u(t- \tau).
\end{equation*}
The shifts for first- and second-order derivatives are defined similarly:
\begin{equation*}
\dot{u} ^+ =   S _+ ( \dot{u}  ) =   \dot{u}  (t + \tau),
\qquad
\dot{u}  ^{-}  =   S_- (  \dot{u} ) =   \dot{u}  (t- \tau),
\end{equation*}
\begin{equation*}
\ddot{u} ^+ =   S _+ ( \ddot{u}  ) =   \ddot{u}  (t + \tau),
\qquad
\ddot{u}  ^{-}  =   S_- (  \ddot{u} ) =   \ddot{u}  (t- \tau).
\end{equation*}
In order to provide a DODE with two delays, the function $F$ is required to satisfy
\begin{equation}  \label{FDODE}
   \left(  {\partial F  \over  \partial {u}^{-} } \right) ^2
+ \left( {\partial F \over  \partial \dot{u}^{-} } \right) ^2
+ \left( {\partial F \over  \partial \ddot{u}^{-} } \right) ^2
{\not\equiv} 0.
\end{equation}

DODEs  (\ref{DODE0}) have to be supplemented by initial conditions. In contrast to the case of ordinary differential equations, which have initial conditions at a point, initial conditions for DODE (\ref{DODE0}) are given on an initial interval of length $2 \tau$, e.g.,
\begin{equation}    \label{initialvalues}
u(t)   = \varphi (t),
\qquad
\dot{u}(t)   = \dot{\varphi} (t),
\qquad
\ddot{u}(t)   = \ddot{\varphi} (t),
\qquad
t \in [ t _0  -  \tau, t _0 +   \tau    ].
\end{equation}
For simplicity, we assume that the function $\varphi(t)$ is twice differentiable on the interval $[t_0 - \tau, t_0  +  \tau  ]$, although this requirement can be relaxed. One of the procedures for solving DODEs either analytically or numerically is called the {\it method of steps}~\cite{bk:Elsgolts1955}.

\subsection{First integrals of  DODEs}

\label{FisrtIttegrals}

A DODE (\ref{DODE0}) may contain a dependent variable and its derivatives at three points $t^+$, $t$, and $t^-$. Two types of conserved quantities can be introduced: differential first integrals and difference first integrals.

\begin{definition}   \label{differential_first_integral_L}
A quantity
\begin{equation}        \label{differential_integral_L}
I  (   t  ^+,   t, t ^-,
 u  ^+,  u,  u  ^-,
 \dot{u}  ^+,   \dot{u}, \dot{u}  ^-     )
\end{equation}
is called a {\it differential first integral} of DODE (\ref{DODE0}) if it holds constant on solutions of the DODE.
\end{definition}

The differential first integral (\ref{differential_integral_L}) satisfies the equation
\begin{equation}
{\DD} ( I )
=   I _{t ^+ }    +  I _t      +   I _{t ^- }
+ I _{u^+ } \dot{u} ^+    + I _u   \dot{u}    + I _{u  ^- }  \dot{u} ^-
+ I _{ \dot{u} ^+ }    \ddot{u} ^+    + I _{  \dot{u} }    \ddot{u}   + I _{  \dot{u} ^- }    \ddot{u} ^-
= 0,
\end{equation}
which should hold for any solution of the considered DODE (\ref{DODE0}).

In addition to the differential first integral, one can define a difference integral.

\begin{definition}   \label{deference_first_integral_L}
Quantity
\begin{equation}
J  (     t, t ^-,
      u,  u  ^-,
     \dot{u}, \dot{u}  ^-,
\ddot{u}, \ddot{u}  ^-       )
\end{equation}
is called a {\it difference first integral} of DODE (\ref{DODE0}) if satisfies the equation
\begin{equation}
( S_+   - 1 )   J  = 0
\end{equation}
on the solutions of the DODE.
\end{definition}

We illustrate the definitions of the first integrals with a simple example.

\begin{example}
The DODE
\begin{equation*}
  \ddot{u} ^+ =     \ddot{u} ^-
\end{equation*}
has the differential first integral
\begin{equation*}
 I =  \dot{u} ^+ -   \dot{u} ^-
\end{equation*}
and the difference first integral
\begin{equation*}
  J  =  \ddot{u}   -    \ddot{u} ^-.
\end{equation*}
\end{example}

The differential first integral is constant on DODE solutions. In contrast, the difference first integral need not be constant on DODE solutions: it can be a periodic function with period $\tau$, where $\tau$ is the delay parameter.

\section{Lie point symmetries and invariant DODEs}

\label{section_Lie_point}

\subsection{Lie point symmetries}

Consider an infinitesimal transformation group
\begin{equation}  \label{Group}
t   \rightarrow   t ^*  = f(t, u, a)
\approx
t + \xi  ( t, u) a,
\qquad
u   \rightarrow  u ^* = g(t,u,a)
\approx
u + \eta  ( t, u) a,
\end{equation}
where $a$  is the group parameter.
Such transformations are represented by a generator in the standard form \cite{bk:Ovsiannikov1978}:
\begin{equation}     \label{operator1_L}
X
=  {\xi(t,u)}{\partial {} \over \partial t}
+ {\eta(t,u)} {\partial {}   \over \partial u} +  \cdots
\end{equation}

For group analysis of second-order DODE with two delays (\ref{DODE0}), generators should be prolonged to all variables included in the DODE: derivatives $\dot{u} $ and $ \ddot{u}$, and variables at shifted points
$(t^{-},u^{-},\dot{u}^-, \ddot{u}^-) $
and
$(t^{+},u^{+},\dot u^{+}, \ddot{u}^+) $.
This leads to
\begin{multline} \label{operator2_L}
{X}
={\xi}{\partial \over \partial t}
+ {\eta} {\partial  \over \partial u}
+ {\zeta} _1  {\partial  \over \partial \dot{u}}
+ {\zeta} _2  {\partial  \over \partial \ddot{u}}
\\
+ {\xi}^- {\partial \over \partial t^- }
+ {\eta}^-  {\partial  \over \partial u^- }
+ {\zeta} _1 ^-   {\partial  \over \partial \dot{u}^- }
+ {\zeta} _2  ^- {\partial  \over \partial \ddot{u}^- }
\\
+ {\xi}^+ {\partial \over \partial t^+ }
+ {\eta}^+  {\partial  \over \partial u^+ }
+ {\zeta} _1 ^+   {\partial  \over \partial \dot{u}^+ }
+ {\zeta} _2  ^+ {\partial  \over \partial \ddot{u}^+ },
\end{multline}
where
\begin{equation*}
\xi   =  \xi(t,u),
\qquad
\eta   =  \eta (t,u),
\end{equation*}
the coefficients
\begin{equation*}
\zeta _1 = \zeta _1  ( t, u,\dot{u} ) =   {\Dt}    ( \eta ) -  \dot{u}   {\Dt}   ( \xi ),
\qquad
\zeta _2 = \zeta _2  ( t, u,\dot{u}, \ddot{u} ) =  {\Dt}    ( \zeta _1 ) -  \ddot{u}  {\Dt}   ( \xi )
\end{equation*}
are found according to the standard prolongation formulas \cite{bk:Ovsiannikov1978,  bk:Ibragimov1983, bk:Olver1986}, and the coefficients
\begin{equation*}
\xi   ^- = S_- ( \xi) =  \xi(t  ^-,u  ^- ),
\qquad
\eta  ^-   = S_-  ( \eta) =  \eta (t ^-,u ^- ),
\end{equation*}
\begin{equation*}
\zeta _1  ^-  = S_- ( \zeta _1)  =   \zeta _1  ( t ^-, u  ^-,\dot{u}  ^-  ),
\qquad
\zeta _2  ^- = S_-   ( \zeta _2)  =     \zeta _2  ( t ^-, u  ^-,\dot{u}  ^-, \ddot{u}  ^- )  ;
\end{equation*}
\begin{equation*}
\xi   ^+ =   S_+ ( \xi )  =   \xi(t  ^+,u  ^+ ),
\qquad
\eta  ^+   = S_+  (  \eta ) = \eta (t ^+,u ^+ ),
\end{equation*}
\begin{equation*}
\zeta _1  ^+  =   S_+ ( \zeta _1 )    = \zeta _1  ( t ^+, u  ^+,\dot{u}  ^+  ),
\qquad
\zeta _2  ^+ =  S_+ ( \zeta _2 )    = \zeta _2  ( t  ^+, u  ^+,\dot{u}  ^+, \ddot{u}  ^+ )
\end{equation*}
are obtained by the left and right shift operators  $ S_- $ and  $ S_+  $, defined in  (\ref{shifts}).
Here, the operator
\begin{equation}      \label{more_differentiation_L}
{\DD}
= \frac{\partial {}}{\partial t}
+\dot{u} \frac{\partial { }}{\partial u}
+ \ddot{u} \frac{\partial}{\partial\dot{u}}
+ \cdots
+ \frac{\partial {}}{\partial t^-}
+\dot{u}^- \frac{\partial { }}{\partial u^-}
+ \ddot{u}^{-}\frac{\partial}{\partial\dot{u}^{-}}
+ \cdots
+ \frac{\partial}{\partial t^{+}}
+\dot{u}^{+}\frac{\partial}{\partial u^{+}}
+\ddot{u}^{+}\frac{\partial}{\partial\dot{u}^{+}}
+ \cdots
\end{equation}
provides the total derivative.

\subsection{The invariance  of DODEs}




To consider invariant DODEs (\ref{DODE_EQ}) with a constant delay (\ref{delays}), we require both equations (\ref{DODE_EQ}) and (\ref{delays}) to be invariant together.

In the present paper, we restrict the function $\xi$ to depend on $t$ only: $\xi=\xi(t)$. In this case, one can single out the invariance condition for system (\ref{DODE_EQ}) and (\ref{delays}), and consider the invariance of the equations separately. The infinitesimal criterion for the invariance of an equation (\ref{DODE_EQ}) becomes
\begin{equation} \label{crit}
\left.
 {X}
(\ddot{u}^+  - F ) \right|_{\ddot{u}^+  =  F,\  t^+-t=t-t^-} =0,
\end{equation}
for the prolonged generators (\ref{operator2_L}).

For an arbitrary $t$, the delay parameter $\tau$ is assumed to have the same value to the right and left of $t$:
\begin{equation}
  t^+ -  {t }= t -  {t ^-} = \tau.
\end{equation}
We need this relation to be preserved under group transformations, i.e.,
\begin{equation*}
(t^+) ^* -  t  ^* =t ^* - ( t ^- ) ^* = \tau ^*  .
\end{equation*}
This leads to the infinitesimal  condition
\begin{equation*}
\xi ( t^+ )  - \xi ( t  ) =
 \xi ( t )  - \xi ( t^-  ).
\end{equation*}
The latter equation has the following solution:
\begin{equation}\label{regGRID}
\xi (t) = f(t) t   + g(t) ,
\end{equation}
where $f(t)$  and   $g(t)$ are  arbitrary periodic functions with the period  $\tau$. Notice that this solution allows the delay parameter to be changed.


The following requirement for a transformation of the delay parameter is that the time scale must remain unchanged after the transformation.

To keep time homogeneity of transformations, we have to preserve the following relation for pairwise distinguished points $t_1,t_2,t_3,t_4$:
\begin{equation}   \label{ratio}
t_1-t_2 = \gamma(t_3-t_4),
\end{equation}
where $\gamma$ is constant.
Applying generator (\ref{operator1_L}), one gets
\begin{equation*}   
 \left( \xi(t_1)- \xi(t_2) -
\gamma(\xi(t_3)-\xi(t_4))\right)|_{(\ref{ratio})}=0.
\end{equation*}
Satisfying  this condition, one obtains the following result:
\begin{equation}
\xi (t) = \alpha t   + \beta   ,\label{cont}
\end{equation}
where $\alpha$ and $\beta$ are arbitrary constants.

Notice that (\ref{cont}) is not required for all results obtained below: some of the requirements can be relaxed.


\section{Lagrangian approach for DODEs}

\label{Lagrangian_framework}

This section describes the Lagrangian approach to delay ordinary differential equations and provides the Noether theorem, established in \cite{DKM2023}.

\subsection{ODE case: the Euler--Lagrange equation}

It is well known~\cite{Gelfand_Fomin,  Abraham, Goldstein, Arnold, Marsden} that variations of the action functional
\begin{equation}   \label{functional_L}
{\cal L}
=
\int_{a}^{b}
L (t,{u},\dot{u})  \ dt,
\end{equation}
where   $L (t,{u},\dot{u})  $ is a Lagrangian function, lead to the Euler-Lagrange equation
\begin{equation}   \label{Euler_Lagrange}
{ \delta {L}   \over \delta u}
=
\frac{\partial L}{\partial u}
- D   \left( \frac{\partial L }{\partial  \dot{u}}  \right)
= 0,
\end{equation}
where
\begin{equation*}
 D
=\frac{\partial}{\partial t}
+\dot{u} \frac{\partial}{\partial u}
+\ddot{u} \frac{\partial}{\partial \dot{u} }
+  \cdots
 \end{equation*}
is the total differentiation operator.
We consider a one-dimensional case $ u \in   \mathbb{R}$ for simplicity.

\begin{remark}   \label{remark_equivalence_L}
In the case of ODEs, the variational equations corresponding to variations of the dependent and independent variables are equivalent. For the first-order Lagrangians $ L =  L ( t, u, \dot{u} )  $, these equations are the Euler–Lagrange equation (\ref{Euler_Lagrange}) and the Du Bois-Reymond equation
\begin{equation}
{ \delta  L \over \delta t}
=
{ \partial    L \over \partial t}
+ {\Dt}  \left( \dot{u}  { \partial  L      \over \partial \dot{u} }   -  L   \right)  =  0,
\end{equation}
respectively.
It is not difficult to check that   these equations are proportional
\begin{equation}     \label{relation}
{ \delta L \over \delta t}
= -    \dot{u}
{ \delta L \over \delta u},
\end{equation}
and, therefore, equivalent.
\end{remark}

\subsection{Variational equations for delay functionals}

\label{local_extremal}

Consider a first-order functional with one constant delay
\begin{subequations}      \label{functional_def_L}
\begin{gather}
{\cal L} = \int_{a}^{b}
{ L}(t, t^-,u,u^-,\dot{u},\dot{u}^-)  dt
\\
t - t^- = \tau,
\qquad
\tau = \mbox{const}.
\end{gather}
\end{subequations}
and the Lagrangian function $L$ satisfying
\begin{equation}
   \left(  {\partial L   \over  \partial {u} } \right) ^2
+ \left( {\partial L  \over  \partial \dot{u} } \right) ^2
{\not\equiv} 0,
\qquad
 \left(  {\partial L   \over  \partial {u}^{-} } \right) ^2
+ \left( {\partial L  \over  \partial \dot{u}^{-} } \right) ^2
{\not\equiv} 0.
\end{equation}

Let the interval $ [ t_1, t_2] $ be such that $a  \leq  t_1< t_2 \leq   b -\tau $.
We apply slight perturbations of the independent and dependent variables given by
\begin{equation}
\label{eq:09mar.1_L}
t _{\varepsilon} = t +
\varphi (t)  \varepsilon,
\qquad
u _{\varepsilon} = u +
\psi (t)  \varepsilon,
\qquad
t_1   \leq t \leq   t_2,
\end{equation}
where $ \varphi (t) $  and $ \psi (t) $ are differentiable functions satisfying
\begin{equation}    \label{boundary}
\varphi (s )  =  0
\quad
\mbox{and}
\quad
\psi (s )  =  0
\qquad
\mbox{for}
\qquad
s \in ( - \infty, t_1  ]  \cup [ t_2, \infty)
\end{equation}
and $ \varepsilon$ is a small parameter.
Such perturbations produce variations of the derivative $ \dot{u} $ and the differential $ dt $. The variational equations were derived in \cite{DKM2023}.

For $\varphi=0$ and $\psi\neq 0$,  we obtain  the extremal delay equation
\begin{equation}   \label{variational_u}
{\deltauL}
=
\frac{\partial { L}}{\partial u}
+ \frac{\partial { L^+}}{\partial u}
- {\DD}  \left( \frac{\partial { L}}{\partial \dot{u}}
 +   \frac{\partial { L^+}}{\partial \dot{u}} \right)
=0,
\end{equation}
which represents the `vertical variation', i.e., a variational equation for the variation of the dependent variable $u$.
This equation is known since Elsgolts  \cite{bk:Elsgolts1955} (see also  \cite{Elsgolts2, Elsgolts3}). We call it the {\it Elsgolts equation}.

For the case $\varphi\neq 0$ and $\psi =0$, we define a `horizontal variation'
\begin{equation}  \label{variational_t_L}
{\frac{\delta  { L} }{\delta t}}
  = {  \partial { L} \over \partial t}
+ { \partial { L} ^+ \over \partial t }
+  {\DD} \left( \dot{u}  { \partial { L} \over \partial \dot{u} }
+ \dot{u}  { \partial { L} ^+ \over \partial \dot{u} }
-  { L} \right)
 = 0.
 \end{equation}

For variations defined by the symmetry (\ref{operator1_L}), we multiply the coefficients of the generator $X$ by a  function $\varphi(t)$ satisfying (\ref{boundary})
\begin{equation}
t _{\varepsilon} = t +
\varphi(t)\xi (t,u)  \varepsilon,
\qquad
u _{\varepsilon} = u +
\varphi(t)\eta  (t,u)  \varepsilon.
\end{equation}
In this case, the variation leads to the equation
\begin{equation}    \label{localextremal_L}
\xi \left[ { \partial { L} \over \partial t}
+ { \partial { L} ^+ \over \partial t }
+  {\DD} \left( \dot{u}  { \partial { L} \over \partial \dot{u} }
+ \dot{u}  { \partial { L} ^+ \over \partial \dot{u} }
- { L} \right) \right]
+\eta \left[
 { \partial { L} \over \partial u}
 +  { \partial { L}^+  \over \partial u }
-  {\DD} \left(  { \partial { L} \over \partial \dot{u} }
+
{ \partial { L} ^+ \over \partial  { \dot{u}  } } \right)
 \right] = 0,
\end{equation}
which depends {\it explicitly} on $\xi$ and $\eta$, i.e., on the considered given group.
It can be rewritten in the form
\begin{equation}         \label{quas_L}
\xi   { \delta L \over \delta t }
+  \eta    {\deltauL}
= 0.
\end{equation}
We call this equation the {\it local extremal equation}. Notice that a local extremal equation is a second-order DODE containing $L$ and the shifted to the right Lagrangian $L^+$, and by virtue of the equation has {\it two delays}. The local extremal equation gives the necessary condition for any Lagrangian to achieve an extremal value for variations along orbits of the considered Lie group.

\begin{remark}
In contrast to the ODE case (see remark \ref{remark_equivalence_L}),  the vertical variational equation  (\ref{variational_u})  and the horizontal variational equation (\ref{variational_t_L}) are not equivalent.
\end{remark}

\subsection{Invariance of delay functionals}

\label{section_Variational}

Invariance of first-order functional with one delay for a one-parameter group of point transformations  (\ref{Group})  was examined in \cite{DKM2023}. It was shown that it is equivalent to the invariance of the elementary action
\begin{equation}\label{Group1_L}
{ L}(t,t^-, u,u^-,\dot{u},\dot{u}^-) dt
={ L}(f,f^-, g,g^-,\dot{g},\dot{g}^- )  {\Dt}  ( f ) dt.
\end{equation}
Thus, the functional is invariant if and only if the elementary action is invariant. It leads to the criterion for invariance of the delay functional.

\begin{theorem}
\label{thm_D1_L} {\bf (Invariance of Lagrangian)}
The functional (\ref{functional_def_L}) is invariant with respect to the group of transformations  with the generator   (\ref{operator1_L}) if and only if
\begin{equation}  \label{Group2_L}
X L + L  {\Dt}  ( \xi  )  = 0.
\end{equation}
\end{theorem}

In detail, the invariance condition (\ref{Group2_L}) states
\begin{equation}   \label{invariance_functional_L}
\xi\frac{\partial {L}}{\partial t}
+ \xi^-\frac{\partial {L}}{\partial {t^{-}}}
+ \eta\frac{\partial {L}}{\partial u}
+ \eta^-   \frac{\partial {L}}{\partial  {u^-}}
+ \zeta    _1   \frac{\partial {L}}{\partial \dot u}
+  \zeta^-   _1  \frac{\partial {L}}{\partial \dot u^-}
+ {L}  {\Dt} (\xi)=0.
\end{equation}

\subsection{Noether's identity and Noether-type theorem}

\label{section_Noether_L}

In this subsection, we introduce the Noether operator identity, which relates the invariance of a Lagrangian (\ref{Group2_L}), a local extremal equation (\ref{localextremal_L}), and conserved quantities.


\begin{lemma}
\label{lem_D3_L}
{\bf (Noether's identity)}
The following identity holds
\begin{equation}   \label{identity1_L}
X L+ L {\Dt}( \xi  )
=
\xi { \delta L \over \delta  t }
+
\eta   {\deltauL}
+
{\DD}  ( C^I )
+
( 1 - S_+ )  C^J,
\end{equation}
where
\begin{equation}        \label{continuous_L}
C
=
\xi {L}
 + ( \eta - \dot{u} \xi )   \left( { \partial {L} \over \partial \dot{u} }
  +   { \partial {L} ^+  \over \partial \dot{u} } \right)
\end{equation}
and
\begin{equation}        \label{difference_L}
P
=
 \xi^-  \frac{\partial {L}}{\partial t^-}
+ \eta^-  \frac{\partial {L}}{\partial u^-}
+  \zeta _1  ^-    \frac{\partial {L}}{\partial \dot{u}^-}.
\end{equation}
\end{lemma}

Direct computations prove the identity. It yields various versions of the Noether theorem for delay differential equations. We recall that invariance of the delay functional (\ref{Group2_L}) does not require invariance of the delay equation (\ref{delays}).

\begin{theorem}  \label{main_Noether_L}
{\bf (Noether's theorem)}
Let a delay functional (\ref{functional_def_L}) be invariant for the group action corresponding to the generator (\ref{operator1_L}) on solutions of the local extremal equation
\begin{equation}    \label{quasi_extremal_L}
\xi { \delta L \over \delta  t }
+
\eta
 {\deltauL}
= 0.
\end{equation}
Then the differential-difference relation
\begin{equation}   \label{dd_L}
  {\DD}  ( C )
=
( S_+ -1  )  P
\end{equation}
holds on solutions of this equation.
\end{theorem}

\noindent {\it Proof.}
The result follows from identity (\ref{identity1_L}).
\hfill $\Box$

\medskip

Theorem \ref{main_Noether_L}  has an extension for the {\it divergence invariant} Lagrangians. Such an extension for the Noether theorem was first proposed in \cite{Bessel_Hagen}.

\begin{corollary}    \label{dd_generalization_L}
Let delay functional  (\ref{functional_def_L}) satisfy the condition
\begin{equation}   \label{dd_invariance_L}
X L+ L  {\Dt}  ( \xi  )    =   {\DD}  ( V  )  +   ( 1  -  S_+  )  W
\end{equation}
with some functions
$
V   (   t  ^+,   t, t ^-,
 u  ^+,  u,  u  ^-,
 \dot{u}  ^+,   \dot{u}, \dot{u}  ^-     )
$
and
$
W  (     t, t ^-,
      u,  u  ^-,
     \dot{u}, \dot{u}  ^-,
\ddot{u}, \ddot{u}  ^-       )
$
on solutions of the local extremal equation (\ref{quasi_extremal_L}).
Then the differential-difference relation
\begin{equation}       \label{dd_divergence_L}
  {\DD}  ( C - V )
=
( S_+  - 1 )  (  P - W  )
\end{equation}
holds on solutions of this equation.
\end{corollary}

Condition (\ref{dd_invariance_L}) means the divergent invariance of the Lagrangian. We call the terms on the right side, namely $ {\DD} ( V ) $ and $ ( 1 - S_+ ) W $,
as differential divergence and difference divergence, respectively.


For some DODE, the differential-difference relation (\ref{dd_L}) can be converted into a differential first integral or a difference first integral.

\begin{corollary}      \label{proposition_differential_L}
If there holds
\begin{equation}
( S_+ - 1  )  P  = {\DD}  ( V )
\end{equation}
with some function
$
V   (   t  ^+,   t, t ^-,
 u  ^+,  u,  u  ^-,
 \dot{u}  ^+,   \dot{u}, \dot{u}  ^-     )
$,
then the differential-difference relation (\ref{dd_L}) provides the differential first integral
\begin{equation}
I = C - V.
\end{equation}
\end{corollary}

\begin{corollary}      \label{proposition_difference_L}
If there holds
\begin{equation}
{\DD}  ( C )  =  ( S_+  - 1 )   W,
\end{equation}
with some function
$
W  (     t, t ^-,
      u,  u  ^-,
     \dot{u}, \dot{u}  ^-,
\ddot{u}, \ddot{u}  ^-       )
$,
then the differential-difference relation (\ref{dd_L}) provides the difference  first integral
\begin{equation}
 J =   P  -  W.
\end{equation}
\end{corollary}

\subsection{Example: delay oscillator}

\label{Linear_oscillator_1_L}

In this example, we illustrate the basic version of Noether's Theorem \ref{main_Noether_L}, its corollary, and how to transform differential-difference relations into differential first integrals.

Consider the Lagrangian function
\begin{equation}         \label{example_1_Lagrangian}
L = \dot{u}  \dot{u} ^-   -      u   u  ^-
\end{equation}
and the symmetries
\begin{equation}
X_1  =  \cos t {  \partial  \over \partial u },
\qquad
X_2  =  \sin t {  \partial  \over \partial u },
\qquad
X_3  =  {  \partial  \over \partial t },
\qquad
X_4  =   u  {  \partial  \over \partial u }.
\end{equation}

The symmetries $ X_1$  and  $ X_2$ are linearly connected. For them,  the local extremal equation  (\ref{quasi_extremal_L}) is the Elsgolts equation
\begin{equation}     \label{example_1_delta_u}
{\deltauL}
=  -     {u} ^{-}  -   {u} ^{+}    -    \ddot{u} ^{-}  -   \ddot{u} ^{+}
= 0.
\end{equation}

The symmetry $  X_1  $ satisfies the divergence invariance condition  (\ref{dd_invariance_L})
\begin{equation*}
{X}_1 L   + L {\Dt} (\xi _1)
=
{\DD} ( -   \sin  t  ^-     {u}    -      \sin  t    {\ } {u} ^{-}    ).
\end{equation*}
Therefore, the differential-difference relation  (\ref{dd_divergence_L}) with
\begin{equation*}
C  _1
=   \cos  t     (   \dot{u} ^{+} +  \dot{u} ^{-} )
+       \sin  t  ^-     {u}   +    \sin  t  {\  }   {u} ^{-},
\qquad
P  _1
=  -  \cos  t  ^-  u  -       \sin  t  ^-   \dot{u}
\end{equation*}
holds on solutions of equation    (\ref{example_1_delta_u}).
Using Corollary  \ref{proposition_differential_L} and
\begin{equation*}
(  S_+ -1 )   P  _1
=  {\DD} (     -       \sin  t  {\ }  {u}  ^+     +        \sin  t  ^-  {u}    ),
\end{equation*}
we find the differential first integral
\begin{equation}
I_1 =  \cos   t     (   \dot{u} ^{+} +  \dot{u} ^{-} )   +   \sin  t  (    {u} ^{+} + {u} ^{-}  ).
\end{equation}

For the symmetry  $ X_2$,  the divergence invariance condition is
\begin{equation*}
{X} _2  L   + L {\Dt} (\xi _2 )
=
{\DD} (  \cos  t  ^-     {u}   +      \cos  t   {\ }   {u} ^{-}    ).
\end{equation*}
The components of the differential-difference relation are
\begin{equation*}
C  _2
=   \sin  t     (   \dot{u} ^{+} +  \dot{u} ^{-} )
-        \cos  t  ^-     {u}  -    \cos  t    {\ }  {u} ^{-},
\qquad
P  _2
=  -  \sin  t  ^-  u  +        \cos  t  ^-   \dot{u}.
\end{equation*}
Using
\begin{equation*}
(  S_+ -1 )   P _2
=  {\DD} (        \cos  t   {\ } {u}  ^+    -       \cos  t  ^-  {u}    ),
\end{equation*}
we transform  the differential-difference relation into the differential first integral
\begin{equation}
I_2 =  \sin   t     (   \dot{u} ^{+} +  \dot{u} ^{-} )  -    \cos t  (    {u} ^{+} + {u} ^{-}  ).
\end{equation}


For the symmetry $X_3 $, we get the horizontal variational equation
\begin{equation}     \label{example_1_delta_t}
{ \delta L \over \delta t}
= {\DD}  (  \dot{u}  \dot{u} ^+  + u  u^- )
=  \ddot{u}  \dot{u} ^+    +      \dot{u}  \ddot{u} ^+      +  \dot{u} u^-    + u   \dot{u} ^-
= 0.
\end{equation}
Notice that equation (\ref{example_1_delta_u}) is linear, while equation (\ref{example_1_delta_t}) is nonlinear.

The symmetry  $ X_3$ is variational, i.e.,  it satisfies
\begin{equation*}
{X}_3  L   + L {\Dt} (\xi _3)
=
0.
\end{equation*}
It leads to the differential-difference relation (\ref{dd_L})
with
\begin{equation*}
C  _3
=    -   \dot{u}  \dot{u} ^+    -    u  u^-,
\qquad
P  _3
\equiv 0.
\end{equation*}
This relation is actually the differential first integral
\begin{equation}
I_3 =  C  _3   =  -    \dot{u}  \dot{u} ^+  -  u  u^-,
\end{equation}
which holds on solutions of the horizontal variational equation (\ref{example_1_delta_t}).

We note that the symmetries $X_1$ and $X_2$ give the first integrals $I_1$ and $I_2$ for the Elsholtz equation. In contrast, the symmetry $X_3$ gives the first integral $I_3$ for the horizontal variational equation (\ref{example_1_delta_t}): the first integral $I_3$ does not hold on solutions of the Elsgolts equation (\ref{example_1_delta_u}).


Variational equations can admit symmetries that are neither variational nor divergence symmetries of the Lagrangians. For example, both equations (\ref{example_1_delta_u}) and  (\ref{example_1_delta_t}) are invariant with respect to the scaling of the dependent variable, which is represented  by the generator $X_4$, while this symmetry is not admitted by the Lagrangian   (\ref{example_1_Lagrangian}):
\begin{equation*}
{X}_4 L   + L {\Dt} (\xi _4)  = 2 L ,
\end{equation*}
where the right-hand side $2 L$  can not be presented as a divergence.

\medskip

The differential first integrals  $I_1$  and $I_2$ can be used to provide solutions of equation (\ref{example_1_delta_u}). Setting them equal to constants
\begin{equation*}
I_1 = A,
\qquad
I_2 = B,
\end{equation*}
we get
\begin{equation*}
u^+  +  u ^-    = A  \sin t    -   B   \cos t.
\end{equation*}

The above general solution can be used to obtain a solution of an initial-value problem by means of just algebraic manipulation. For convenience, we rewrite this relation in the shifted form
\begin{equation}   \label{recursion_L}
u ( t)  +  u  (t - 2 \tau )     = A  \sin (  t - \tau )     -   B   \cos ( t - \tau ).
\end{equation}
We start with the initial values
\begin{equation*}
u  (t)   =
\varphi (t),
\qquad
t \in  [ -  2 \tau, 0  ].
\end{equation*}
For simplicity, we assume that the function $\varphi(t)$ has a continuous first derivative on this interval. The initial conditions give
\begin{equation*}
 A = \dot{\varphi} ( 0 ) +  \dot{\varphi} ( - 2 \tau ),
\qquad
B  =   -    {\varphi} ( 0 )   -   {\varphi} ( - 2 \tau ).
\end{equation*}

Using (\ref{recursion_L}), we obtain
\begin{equation*}
u  (t)
= A  \sin (  t - \tau )       -  B \cos    (  t - \tau )    -  \varphi (t -2 \tau ),
\qquad
t \in [ 0,  2 \tau ]   ;
\end{equation*}
\begin{multline*}
u (t)
= A  \sin (  t - \tau )     -  B  \cos    (  t - \tau )   -  u    (t -2 \tau )
\\
= A  \sin (  t - \tau )     -  B  \cos    (  t - \tau )
-  A  \sin (  t - 3 \tau )     +   B  \cos    (  t - 3 \tau )
     +   \varphi (t - 4 \tau ),
\qquad
t \in [  2 \tau,  4 \tau  ]   ;
\end{multline*}
and so on.
By virtue of these relations, one can find the solution $ u(t) $, $ t \in [ \tau, \infty )$ recursively, starting from the initial data. In contrast to the method of steps~\cite{bk:Elsgolts1955}, this recursive procedure does not require any integration.


\section{Hamiltonian approach for DODEs}

\label{Hamiltonian_framework}

The Hamiltonian approach to delay ordinary differential equations was developed in \cite{DKM2025}. It is related to the Lagrangian approach via a delay analog of the Legendre transformation.

\subsection{ODE case: the canonical Hamiltonian equations}

We shortly overview the variational approach for canonical Hamiltonian ODEs in the case of scalar dependent functions ${q (t) }$  and  $ {p (t) }$. Similarly, we will construct a Hamiltonian approach for DODEs.

We consider the canonical Hamiltonian equations~\cite{Goldstein, Arnold}
\begin{equation}    \label{canonical}
\dot{q}=\frac{\partial H}{\partial{p}},
\qquad
\dot{p}=-\frac{\partial H}{\partial{q}},
\end{equation}
for some Hamiltonian function $H=H(t,{q},{p})$.
These equations can be obtained by the variational principle in the phase space $({q},{p})$ from the action functional
\begin{equation}   \label{functional_H}
{\cal H}
=
\int_{a}^{b} \left(  p \dot{q} - H(t,{q},{p}) \right) d t
\end{equation}
(see, for example, \cite{Gelfand_Fomin}).

Let us note that the canonical Hamiltonian equations~(\ref{canonical}) can be obtained by the action of the variational operators
\begin{equation*}   
\frac{\delta}{\delta p}
=
\frac{\partial}{\partial p}
- D\frac{\partial}{\partial\dot{p}},
\qquad
\frac{\delta}{\delta q}
=
\frac{\partial}{\partial q}
-D \frac{\partial}{\partial\dot{q}},
\end{equation*}
where $D$ is the operator of total differentiation with respect to time
\begin{equation*}  
 D
=\frac{\partial}{\partial t}
+\dot{q}\frac{\partial}{\partial q}
+\dot{p}\frac{\partial}{\partial p}+  \cdots,
 \end{equation*}
on the integrand of the functional~(\ref{functional_H}), namely on the function
\begin{equation}    \label{integrand}
\tilde{H}  =  p \dot{q} - H(t, q,{p}).
\end{equation}
In detail, we obtain
\begin{equation}   \label{varoperator2}
\frac{\delta  \tilde{H}   }{\delta p}
=
\dot{q}
-  \frac{\partial H} {\partial  p }
=  0,
\qquad
\frac{\delta  \tilde{H}   }{\delta q}
=
- \dot{p}
-  \frac{\partial H} {\partial  q }
= 0.
\end{equation}

\begin{remark}   \label{the_third_equation}
We can also consider the variation with respect to the independent variable $t$
\begin{equation}     \label{varoperator3}
{\delta  \tilde{H}   \over \delta t }
=
D(H) - \frac{\partial H}{\partial t}
=0.
\end{equation}
This equation holds on the solutions of the canonical Hamiltonian equations~(\ref{canonical}):
\begin{equation*}  
 \left.D(H)\right|_{\dot{q}=H_{p},\
\dot{p}=-H_{q}}
 =\left[
\frac{\partial H}{\partial t}
+\dot{q}\frac{\partial H}{\partial q}
 +\dot{p}\frac{\partial H}{\partial p}
\right]_{\dot{q}=H_{p}, \ \dot{p}=-H_{q}}
=\frac{\partial H}{\partial t}.
 \end{equation*}
\end{remark}

The relationship of the Hamiltonian function  $ H(t,{q},{p})$ with  the Lagrangian function $L(t,{q},{\dot{q}})$ is given by the Legendre transformation
\begin{equation}  \label{Legendre}
H (t,{q},{p})
= p \dot{q}
- L (t,{q},{\dot{q}}),
\end{equation}
where we should substitute $ \dot{q} $  expressed from
\begin{equation}  \label{to_find_dot_q}
{p}=\frac{\partial L}{\partial\dot{q}}  (t,{q},{\dot{q}})
\end{equation}
in the right-hand side~\cite{Gelfand_Fomin}.
Equation~(\ref{to_find_dot_q}) can be resolved for  $ \dot{q} $ if  $ \frac{\partial^2 L}{\partial  \dot{q} ^2}   \neq 0 $.

The Legendre relation~(\ref{Legendre}) makes it possible to establish equivalence of Euler--Lagrange equation~(\ref{Euler_Lagrange}) and canonical Hamiltonian equations~(\ref{canonical}) (see, for example,~\cite{Arnold}). In addition to the relation~(\ref{to_find_dot_q}), there hold
\begin{equation}
\dot{q}  ={\partial H \over \partial p},
\qquad
{\partial H \over \partial q} = -   {\partial L \over \partial q},
 \qquad
{\partial H \over \partial t}  = -   {\partial L \over \partial t}.
\end{equation}

Lie point symmetries of canonical Hamiltonian equations~(\ref{canonical}) are given by the generators of the form
\begin{equation}  \label{symmetry_H}
X
= \xi(t,{q},{p})\frac{\partial}{\partial t}
+ \eta(t,{q},{p})\frac{\partial}{\partial q}
+ \nu(t,{q},{p})\frac{\partial}{\partial p},
\end{equation}
which are prolonged to the derivatives according to the standard prolongation formulas~\cite{bk:Ovsiannikov1978, bk:Ibragimov1983, bk:Olver1986}.

\begin{remark}
It should be noticed that the Legendre transformation~(\ref{Legendre}) is not a point transformation. Hence, there is no one-to-one correspondence between Lie point symmetries of the Euler-Lagrange equation~(\ref{Euler_Lagrange}), which are given by the generators of the form~(\ref{operator1_L}), and Lie point symmetries of the canonical Hamiltonian equations, which are presented by the generators of the form~(\ref{symmetry_H}).
\end{remark}

\subsection{Variational equations for delay functionals}

\label{section_delay_ODEs}

Similarly to the continuous case, one can consider a delay analog of the functional~(\ref{functional_H}). We introduce the Hamiltonian functional with one delay

\begin{subequations}
\begin{gather}
\label{functional_H_delay}
 {\cal H}
=
\int  _a ^b
p^{-}(\alpha_{1} d \dot{q}+\alpha_{2} d \dot{q}^{-})
+p(\alpha_{3} d \dot{q}+\alpha_{4} d \dot{q}^{-})
-H(t,t^{-},q,q^{-},p,p^{-}) \ d t,
\\
\label{delay_H}
t - t^- = \tau,
\qquad
\tau = \mbox{const}.
\end{gather}
\end{subequations}
where   $ \alpha_{i} $, $ i = 1,2,3,4$ are some constants.
One can consider more complicated cases of the coefficients $ \alpha_{i} $. {Non}constant $ \alpha_{i} $ are treated in~\cite{DKM2024arXiv}.


For the variations of the dependent variables $p$  and $q$ and the independent variable $t$ taken on the interval $ [ a. b - \tau] $, we obtain equations
\begin{subequations}        \label{delay_variation}
\begin{gather}
 \label{delay_variation_p}
{ \delta  \tilde{H}  \over \delta p }
=
\alpha_{1}\dot{q}^{+}
+(\alpha_{2}+\alpha_{3})\dot{q}
+\alpha_{4}\dot{q}^{-}
-\frac{\partial }{\partial{p}}   (H+H^{+})
= 0,
\\
 \label{delay_variation_q}
{ \delta  \tilde{H}  \over \delta q }
= - \left(
\alpha_{4}\dot{p}^{+}
+(\alpha_{2}+\alpha_{3})\dot{p}
+\alpha_{1}\dot{p}^{-}
+\frac{\partial  }{\partial q} (H+H^{+})
\right)
=  0,
\\
  \label{delay_variation_t}
{ \delta  \tilde{H}  \over \delta t }
=
    D  [\alpha_{2}(p\dot{q}-p^{-}\dot{q}^{-})+\alpha_{4}(p^{+}\dot{q}-p\dot{q}^{-}) ]
    +D(H)
   - \frac{\partial  }{\partial t}   (H+H^{+})
= 0,
\end{gather}
\end{subequations}
where
\begin{equation}       \label{H_tilde}
\tilde{H}
=
p^{-}(\alpha_{1}\dot{q}+\alpha_{2}\dot{q}^{-})
+p(\alpha_{3}\dot{q}+\alpha_{4}\dot{q}^{-})
-H(t,t^{-},q,q^{-},p,p^{-})
\end{equation}
is the function corresponding to the integrand of functional~(\ref{functional_H_delay}).
Here, the operator of total differentiation   $D$  is
\begin{equation*}  
 D
=\frac{\partial}{\partial t}
+\dot{q}\frac{\partial}{\partial q}
+\dot{p}\frac{\partial}{\partial p}
+  \cdots
+ \frac{\partial}{\partial t^-}
+\dot{q}^- \frac{\partial}{\partial q^-}
+\dot{p}^- \frac{\partial}{\partial p^-}
+  \cdots
+ \frac{\partial}{\partial t^+}
+\dot{q}^+ \frac{\partial}{\partial q^+}
+\dot{p}^+ \frac{\partial}{\partial p^+}
+  \cdots
 \end{equation*}
The first and second operators perform the 'vertical variations' (i.e., the variations with respect to the dependent variables $p$ and $q$), the last one takes the 'horizontal variation' (the variation with respect to the independent variable $t$).

\begin{remark}
In contrast to the canonical Hamiltonian ODEs (see Remark~\ref{the_third_equation}), the equation~(\ref{delay_variation_t}) does not hold on the solutions of the equations~(\ref{delay_variation_p})  and~(\ref{delay_variation_q}), supplemented by the delay equation (\ref{delays}).
\end{remark}

For the variations along group orbits corresponding to the generator~(\ref{symmetry_H}), we obtain the variational equation
\begin{equation}       \label{quasi_H}
\xi
{ \delta  \tilde{H}  \over \delta t }
+
\eta
{ \delta  \tilde{H}  \over \delta q }
+ \nu
{ \delta  \tilde{H}  \over \delta p }
= 0.
\end{equation}
This local extremal equation is the Hamiltonian analog of~(\ref{quas_L}).

We  consider the system of equations
\begin{subequations}         \label{delay_canonical}
\begin{gather}    \label{delay_canonical_p}
\alpha_{1}\dot{q}^{+}
+(\alpha_{2}+\alpha_{3})\dot{q}
+\alpha_{4}\dot{q}^{-}
=
\frac{\partial }{\partial{p}}   (H+H^{+}),
\\
     \label{delay_canonical_q}
\alpha_{4}\dot{p}^{+}
+(\alpha_{2}+\alpha_{3})\dot{p}
+\alpha_{1}\dot{p}^{-}
=
-  \frac{\partial }{\partial q}  (H+H^{+}),
\end{gather}
\end{subequations}
given by~(\ref{delay_variation_p}) and~(\ref{delay_variation_q}) as the {\it delay  canonical Hamiltonian equations}, i.e., the delay analog of the canonical Hamiltonian equations~(\ref{canonical}).
Shortly, they can be presented as
\begin{equation}    \label{delay_canonical_short}
{ \delta  \tilde{H} \over \delta p }  =0,
\qquad
{ \delta  \tilde{H}  \over \delta q }  = 0.
\end{equation}
These equations are first-order DODEs with two delays. They are supplemented by the delay equation~(\ref{delays}). Note that all three variational equations  (\ref{delay_variation}) supplemented by the delay equation~(\ref{delays}) form an overdetermined system.

\subsection{Compatibility of Hamiltonian approach with Lagrangian approach}
\label{section_Hamiltonian_formalism}

In the classical Hamiltonian theory (see, for example,~\cite{Gelfand_Fomin}), the variables $q$ and $p$ are named canonical if the system of equations for them is equivalent to the Euler-Lagrange variational equation. We are following the same strategy for the delay Hamiltonian equations and considering the transition from the Lagrangian to the Hamiltonian approach via a delay version of the Legendre transformation.

\subsubsection{Delay Legendre transformation}

We consider the delay analog of the Legendre transformation
\begin{equation}    \label{delay_Legendre}
H(t,t^{-},q,q^{-},p,p^{-})
=
p^{-}(\alpha_{1}\dot{q}+\alpha_{2}\dot{q}^{-})
+p(\alpha_{3}\dot{q}+\alpha_{4}\dot{q}^{-})
-L  (t,t^{-},q,q^{-},\dot{q},\dot{q}^{-}),
\end{equation}
where $\dot{q} $  and  $\dot{q} ^- $ in the right-hand side are to be substituted as found from
\begin{equation}    \label{234gf}
 \alpha_{1}  {p^- }   +  \alpha_{3}  {p}
 =\frac{\partial L}{\partial\dot{q}},
\qquad
  \alpha_{2}  {p^- }   +  \alpha_{4}  {p}
 =\frac{\partial L}{\partial\dot{q}^{-}}.
\end{equation}
We also obtain relations
\begin{equation}
 \label{conditions_2}
\alpha_{3}\dot{q}+\alpha_{4}\dot{q}^{-}
= \frac{\partial H}{\partial p },
\qquad
\alpha_{1}\dot{q}+\alpha_{2}\dot{q}^{-}
= \frac{\partial H}{\partial p ^- },
\end{equation}
and
\begin{equation}
 \label{conditions_3}
\frac{\partial H}{\partial q }   =  -  \frac{\partial L}{\partial q },
\qquad
\frac{\partial H}{\partial q ^- }   =  -  \frac{\partial L}{\partial q^-  },
\qquad
\frac{\partial H}{\partial t }   =  -  \frac{\partial L}{\partial t },
\qquad
\frac{\partial H}{\partial t^-  }   =  -  \frac{\partial L}{\partial t^- }.
\end{equation}

We are interested in delay canonical Hamiltonian equations~(\ref{delay_canonical}), which correspond to Elsgolts equation~(\ref{variational_u}). Starting with a delay Lagrangian $L$, we obtain the corresponding delay Hamiltonian $H$. The delay Legendre transformation~(\ref{delay_Legendre}) gives the correspondence between the Lagrangian and the Hamiltonian. We impose a {\it compatibility condition}\footnote{Imposing this condition, we avoid contradiction with the Legendre relations in a given point with that in a shifted point, and avoid introducing a double set of variables $p$ as it was done in~\cite{bk:Zhang}.}, which requires that the variables $p$ and $p^- $ expressed from the  Legendre transformation via the variables $\dot{q}$ and  $\dot{q}^-$ satisfy
\begin{equation}    \label{requirement_p}
p=S_{+} (p^- ).
\end{equation}

\subsubsection{Delay Legendre transformation with compatibility condition}

We consider Lagrangians to be quadratic in the derivatives. Namely, the Lagrangians of the form
\begin{equation}     \label{quadratic_L}
{L}
=  { \alpha  \over 2 }    {\dot{q}}^{2}
+  \beta{\dot{q}\dot{q}}^{-}
+  { \gamma    \over 2 } (\dot{q}^{-})^{2}
-   \phi(q,q^{-}),
\end{equation}
where $\alpha$, $\beta \neq 0$, and $\gamma$ are some constants (a more general form of Lagrangians is examined in~\cite{DKM2024arXiv}).
The Elsgolts variational equation for  this Lagrangian takes the form
\begin{equation}     \label{Elsgolts_variational_equation}
{\delta {L} \over \delta q }
=     -  \left(
\beta{\ddot{q}}^{+}
 +  (\alpha+\gamma)\ddot{q}
 +   \beta{\ddot{q}}^{-}
 +   \frac{\partial}{\partial q}     (\phi+\phi^{+})
\right)
=0.
\end{equation}

It was shown in \cite{DKM2025} that the  compatibility condition (\ref{requirement_p}) holds for coefficient
\begin{equation}    \label{coefficients}
\alpha_{2}=\frac{\gamma}{\beta}  \alpha_{1},
\qquad
\alpha_{3}=\frac{\alpha}{\beta}   \alpha_{1},
\qquad
\alpha_{4}=\alpha_{1} \neq 0.
\end{equation}
For these  coefficients, we obtain
\begin{equation}     \label{relations_p}
p =   {\beta \over \alpha_1 }    \dot{q},
\qquad
p^- =    {\beta \over \alpha_1 }    \dot{q}^-.
\end{equation}
The Hamiltonian given by~(\ref{delay_Legendre}) takes the form
\begin{equation}  \label{Hamiltonian}
H=\frac{\alpha_{1}^{2}}{\beta^{2}}
\left(  { \alpha  \over 2 } p{}^{2}  +\beta pp^{-}+ { \gamma \over 2 }  (p^{-}) ^{2} \right)
+ \phi ( q, q^- ).
\end{equation}
For this Hamiltonian, the  delay Hamiltonian equations~(\ref{delay_canonical}) with the  coefficients~(\ref{coefficients}) can be presented as
\begin{subequations}     \label{eq:GenHamEq_d}
\begin{gather}
{\displaystyle
\alpha_{1}\left(\dot{q}^{+}+ \frac{\alpha + \gamma }{\beta}\dot{q}+\dot{q}^{-}\right)
= \frac{\alpha_{1}^{2}}{\beta}
\left(   p^{+}  +    { \alpha    + \gamma  \over \beta} p  +   p^{-}   \right)  },
\\
{\displaystyle  \alpha_{1}\left(\dot{p}^{+}+ \frac{\alpha+ \gamma}{\beta}\dot{p}+\dot{p}^{-}\right)
=-\frac{\partial}{\partial q} }  ( \phi  + \phi ^{+} ).
\end{gather}
\end{subequations}
They provide a decomposition of the second-order Elsgolts equation~(\ref{Elsgolts_variational_equation}) into two first-order DODEs with two delays. A convenient choice  $ \alpha _1 = \beta $ simplifies the relations~(\ref{relations_p}), the Hamiltonian~(\ref{Hamiltonian}), and the delay canonical Hamiltonian equations~(\ref{eq:GenHamEq_d}).

\subsection{Invariance of delay functionals}
\label{section_symmetries}

In section \ref{section_Lie_point}, we introduced Lie point symmetries for the Lagrangian approach. Here, we modify them to the Hamiltonian case. Let group transformations in the space $(t,{q},{p})$ be provided by generators of the form~(\ref{symmetry_H}). To apply the generator~(\ref{symmetry_H})  to delay Hamiltonian equations, we prolong it to all variables involved, namely derivatives $\dot{q} $ and $ \dot{p}$, and variables at the shifted points
$(t^{-},q^{-},  p^-, \dot{q}^-, \dot{p}^-) $
and
$(t^{+},q^{+}, p^+, \dot q^{+}, \dot{p}^+) $.
We obtain the prolonged operator
\begin{multline} \label{operator2}
{X}
={\xi}{\partial \over \partial t}
+ {\eta} {\partial  \over \partial q}
 + \nu  \frac{\partial}{\partial p}
+ {\zeta} _{\eta}   {\partial  \over \partial \dot{q}}
+ {\zeta} _{\nu}   {\partial  \over \partial \dot{p}}
\\
+ {\xi}^- {\partial \over \partial t^- }
+ {\eta}^-  {\partial  \over \partial q^- }
+ \nu ^-  \frac{\partial}{\partial p^-}
+ {\zeta} _{\eta}  ^-   {\partial  \over \partial \dot{q}^- }
+ {\zeta} _{\nu}  ^- {\partial  \over \partial \dot{p}^- }
\\
+ {\xi}^+ {\partial \over \partial t^+ }
+ {\eta}^+  {\partial  \over \partial q^+ }
+ \nu  ^+ \frac{\partial}{\partial p^+}
+ {\zeta} _{\eta}  ^+   {\partial  \over \partial \dot{q}^+ }
+ {\zeta} _{\nu}   ^+  {\partial  \over \partial \dot{p}^+ },
\end{multline}
where
\begin{equation*}
 \xi   =  \xi(t,q,p),
 \qquad
 \eta   =  \eta (t,q,p),
 \qquad
 \nu =  \nu  (t,q,p).
\end{equation*}
The coefficients
\begin{equation*}
{\zeta} _{\eta}   = {\zeta} _{\eta}   ( t, q, p,  \dot{q},  \dot{p} )
=   {\DD}    ( \eta ) -  \dot{q}   {\DD}  ( \xi ),
\qquad
{\zeta} _{\nu}   =  {\zeta} _{\nu}   ( t, q, p,  \dot{q},  \dot{p} )
=   {\DD}    ( \nu  ) -  \dot{p}   {\DD}   ( \xi ),
\end{equation*}
are found according to the standard prolongation formulas~\cite{bk:Ovsiannikov1978, bk:Ibragimov1983, bk:Olver1986}.
The other coefficients are obtained by the left and right shift operators  $ S_- $ and  $ S_+  $:
\begin{equation*}
\xi   ^- = S _- ( \xi),
\quad
\eta  ^-   = S_-  ( \eta),
\quad
\nu ^-   = S_-  ( \nu ),
\quad
{\zeta} _{\eta}   ^-    = S_-  ( {\zeta} _{\eta}  ),
\quad
{\zeta} _{\nu}   ^-   = S_-  (  {\zeta} _{\nu}  ),
\end{equation*}
\begin{equation*}
\xi   ^+ = S _+ ( \xi),
\quad
\eta  ^+  = S_+ ( \eta),
\quad
\nu ^+   = S_+ ( \nu ),
\quad
{\zeta} _{\eta}    ^+   = S_+  ( {\zeta} _{\eta} ),
\quad
{\zeta} _{\nu}   ^+    = S_+ ( {\zeta} _{\nu}  ).
\end{equation*}
The left and right shift operators  $S_{-}  $ and   $S_{+}  $ are defined in~(\ref{shifts}).

As in the Lagrangian approach, the invariance of a delay functional is equivalent to the invariance of the corresponding elementary action. We go directly to the result.

\begin{theorem}   \label{thm_D1_H}
{\bf (Invariance of delay Hamiltonian) }
The functional~(\ref{functional_H}) is invariant with respect to the group transformations  with the generator~(\ref{symmetry_H}) if and only if
\begin{multline}  \label{Group2}
X ( \tilde{H} ) + \tilde{H}   D ( \xi )
\\
= \nu^{-}    (\alpha_{1}\dot{q}+\alpha_{2}\dot{q}^{-})
+   p^{-}   (\alpha_{1}D(\eta)+\alpha_{2}D(\eta^{-}))
+  \nu   (\alpha_{3}\dot{q}+\alpha_{4}\dot{q}^{-})
+  p   (\alpha_{3}D(\eta)+\alpha_{4}D(\eta^{-}))
\\
+(\alpha_{2}p^{-}+\alpha_{4}p)   \dot{q}^{-}   D(\xi-\xi^{-})
-\xi\frac{\partial H}{\partial t}
-\eta\frac{\partial H}{\partial q}
-\nu\frac{\partial H}{\partial p}
-\xi^{-}\frac{\partial H}{\partial t^{-}}
-\eta^{-}\frac{\partial H}{\partial q^{-}}
-\nu^{-}\frac{\partial H}{\partial p^{-}}
\\
-H D(\xi)
= 0.
\end{multline}
\end{theorem}

\subsection{First integrals of delay Hamiltonian equations}
\label{section_first_integrals}

We adopt the form of first integrals, introduced in subsection \ref{FisrtIttegrals}, to delay canonical Hamiltonian equations. The delay Hamiltonian equations can contain the dependent variables at three points: $t^+$, $t$, and $t^{-}$. There can be two types of conserved quantities: differential first integrals and difference first integrals.

\begin{definition}   \label{differential_first_integral_H}
Quantity
\begin{equation}   \label{differential_integral_H}
I(t^{+},t,t^{-},q^{+},q,q^{-},p^{+},p,p^{-})
\end{equation}
is called \textit{a differential first integral} of the delay Hamiltonian equations if it is constant on the solutions of the delay Hamiltonian equations.
\end{definition}

We require that the differential first integrals~(\ref{differential_integral_H}) satisfy the equation
\begin{equation}
D(I)=I_{t^{+}}+I_{t}+I_{t^{-}}+I_{q^{+}}\dot{q}^{+}+I_{q}\dot{q}+I_{q^{-}}\dot{q}^{-}+I_{p^{+}}\dot{p}^{+}+I_{\dot{p}}\dot{p}+I_{\dot{p}^{-}}\dot{p}^{-}=0,
\end{equation}
which should hold on the solutions of the considered delay Hamiltonian equations.

In addition to differential first integrals, we can define difference first integrals.

\begin{definition} \label{deference_first_integral}
Quantity
\begin{equation}
J(t,t^{-},q,q^{-},p,p^{-},\dot{q},\dot{q}^{-},\dot{p},\dot{p}^{-})
\end{equation}
is called \textit{a difference first integral} of the delay Hamiltonian equations if it satisfies the equation
\begin{equation}
(S_{+}-1)J=0
\end{equation}
on the solutions of the delay Hamiltonian equations.
\end{definition}

\subsection{Hamiltonian identity and Noether-type theorem}
\label{section_Noether_H}

Noether-type theorems are formulated based on Hamiltonian identity, which is a Hamiltonian version of the Noether operator identity used in the Lagrangian approach.


\begin{lemma}  \label{lem_D3_H}
{\bf (Hamiltonian identity)}
The following identity holds
\begin{equation}   \label{DelayNoether_b}
X ( \tilde{H} ) + \tilde{H}   D ( \xi )
\equiv
\xi
{ \delta  \tilde{H}  \over \delta t }
+
\eta
{ \delta  \tilde{H}  \over \delta q }
+ \nu
{ \delta  \tilde{H}  \over \delta p }
+  D  ( C )
+ (1 -  S_{+} )    P,
\end{equation}
where
the variations
$ { \delta  \tilde{H}  \over \delta t }  $,
$  { \delta  \tilde{H}  \over \delta q }   $,
and
$  { \delta  \tilde{H}  \over \delta p }   $
are given in~(\ref{delay_variation}),
\begin{equation*}
C
=
\eta( \alpha_{4}{p}^{+}
+(\alpha_{2}+\alpha_{3}){p}
+ \alpha_{1}{p}^{-} )
- \xi(
\alpha_{2}  (  p  \dot{q} - p^{-} \dot{q}^{-}  )
+\alpha_{4} ( p^{+}\dot{q} - p \dot{q}^{-}  )
+ H),
\end{equation*}
and
\begin{equation*}
P
=
(\alpha_{2}p^{-}+\alpha_{4}p)D(\eta^{-})
+ \nu^{-}(\alpha_{1}\dot{q}+\alpha_{2}\dot{q}^{-})
- (\alpha_{2}p^{-}+\alpha_{4}p)\dot{q}^{-}D(\xi^{-})
-   \xi^{-}   \frac{\partial H}{\partial t^{-}}
-  \eta^{-} \frac{\partial H}{\partial q^{-}}
-  \nu^{-} \frac{\partial H}{\partial p^{-}}.
\end{equation*}
\end{lemma}



Based on the Hamiltonian identity, the most general Noether-type theorem is formulated as follows.

\begin{theorem}    \label{Noether_theorem_H}
{\bf (Noether's theorem)}
Let a Hamiltonian with some given constants $\alpha_{1}$, $\alpha_{2}$, $\alpha_{3}$ and  $\alpha_{4}$ be invariant with respect to a one-parameter group of transformations with a generator~(\ref{symmetry_H}), i.e. condition~(\ref{Group2}) hold.
Then,  on the solutions of the local extremal equation
\begin{equation}      \label{quasi_H_b}
\xi
{ \delta  \tilde{H}  \over \delta t }
+
\eta
{ \delta  \tilde{H}  \over \delta q }
+ \nu
{ \delta  \tilde{H}  \over \delta p }
= 0
\end{equation}
there holds the differential-difference relation
\begin{equation}   \label{dd_H}
D  (C)
=  (S_{+}-1)   P.
\end{equation}
\end{theorem}

\noindent {\it Proof.}
The proof follows from the Hamiltonian identity~(\ref{DelayNoether_b}).
\hfill $\Box$

\medskip

For Noether's theorem, it is convenient to generalize the invariance of a delay  Hamiltonian~(\ref{Group2}) and consider {\it divergence invariance}~\cite{Bessel_Hagen}.

\begin{definition}   \label{divergence_invariance}
The delay Hamiltonian  $ H ( t,t^{-},q,q^{-}, p,  p ^{-})  $ is called  divergence invariant for a generator~(\ref{symmetry_H}) if instead of the condition~(\ref{Group2}) it satisfies the condition
\begin{equation}   \label{dd_invariance}
X ( \tilde{H} ) + \tilde{H}   D ( \xi ) =   {\DD}  ( V  )  +   ( 1  -  S_+  )  W
\end{equation}
with some functions
$
V   (   t  ^+,   t, t ^-,
 q  ^+,  q,  q  ^-,
p  ^+,   p, p ^-     )
$
and
$
W  (     t, t ^-,
      q,  q  ^-,  p,  p ^{-},
     \dot{q}, \dot{q}  ^-,
\dot{p}, \dot{p} ^-       )
$
on solutions of local extremal equation~(\ref{quasi_H_b}).
\end{definition}

\begin{corollary}    \label{dd_generalization}
Let delay Hamiltonian functional~(\ref{functional_H_delay}) satisfy divergence invariance~(\ref{dd_invariance}). Then,  on solutions of the local extremal  equation~(\ref{quasi_H_b}) there holds  the differential-difference relation
\begin{equation}       \label{dd_divergence}
  {\DD}  ( C - V )
=
( S_+  - 1 )  (  P - W  ).
\end{equation}
\end{corollary}



Based on the Noether Theorem~\ref{Noether_theorem_H}, it is possible to provide specified results.

\begin{corollary}  \label{proposition_differential}
If the difference in the  differential-difference relation~(\ref{dd_H}) can be presented as a total derivative, i.e.
\begin{equation}
( S_+ - 1  )  P  = {\DD}  ( V )
\end{equation}
with some function $ V   (   t  ^+,   t, t ^-,
 q  ^+,  q,  q  ^-,
 p  ^+,   p, p  ^-     )
$,
then there holds the differential first integral
\begin{equation}
I = C - V.
\end{equation}
\end{corollary}

\begin{corollary}   \label{proposition_difference}
If the total derivative  in the  differential-difference relation~(\ref{dd_H}) can be presented as a difference, i.e.
\begin{equation}
{\DD}  ( C )  =  ( S_+  - 1 )   W,
\end{equation}
with some function
$
W  (     t, t ^-,        q,  q  ^-,    p, p  ^-,
     \dot{q}, \dot{q}  ^-,      \dot{q}, \dot{q}  ^-     )
$,
then, there is the difference first integral
\begin{equation}
 J =   P  -  W.
\end{equation}
\end{corollary}

\subsubsection{Delay canonical Hamiltonian equations}

The system of local extremal equation~(\ref{quasi_H}) and delay parameter equation~(\ref{delays}) is underdetermined. There are several ways to choose a determined system. The most important choice is to consider the delay canonical Hamiltonian equations.

\begin{proposition}
The results of the Theorem~\ref{Noether_theorem_H}  hold for symmetries~(\ref{symmetry_H}) with   $ \xi   \equiv 0 $ and the delay canonical Hamiltonian equations~(\ref{delay_canonical})
with constant delay~(\ref{delays}).
\end{proposition}

\subsection{Example: delay oscillator}
\label{Section_Examples}

We consider the Lagrangian
\begin{equation}
{L} (t,t^{-},q,q^{-},\dot{q},\dot{q}^{-})=\dot{q}\dot{q}^{-}-qq^{-}.
\end{equation}
It  yields the Elsgolts equation
\begin{equation}    \label{Elsgolts_eq_example_1}
\frac{\delta {L} }{\delta q}
=-(\ddot{q}^{+}+\ddot{q}^{-}+q^{+}+q^{-})=0.
\end{equation}

Following~(\ref{coefficients}), we take
\begin{equation}   \label{example_coefficients}
\alpha_1 = 1,
\qquad
\alpha_2 = 0,
\qquad
\alpha_3 = 0,
\qquad
\alpha_4 = 1.
\end{equation}
For these coefficients, the relations~(\ref{relations_p}) take the form
\begin{equation*}
p = \dot{q},
\qquad
 p ^- = \dot{q} ^-.
\end{equation*}

We obtain the Hamiltonian function
\begin{equation}        \label{example_Hamiltonian}
H
= p^{-}\dot{q}+p\dot{q}^{-}-{L}
= p^{-}\dot{q}+p\dot{q}^{-}-\dot{q}\dot{q}^{-}+qq^{-}
=pp^{-}+qq^{-}
\end{equation}
and the delay canonical Hamiltonian equations
\begin{subequations}    \label{HamEq32}
\begin{gather}
\dot{q}^{+}+\dot{q}^{-}
=p^{+}+p^{-},
\\
\dot{p}^{+}+\dot{p}^{-}
=- q^{+} - q^{-}.
\end{gather}
\end{subequations}
We consider these equations with the delay equation~(\ref{delays}). One can easily verify that these Hamiltonian equations are equivalent to the Elsgolts equation~(\ref{Elsgolts_eq_example_1}).


\medskip

For Hamiltonian   (\ref{example_Hamiltonian}) and coefficients  (\ref{example_coefficients}), we obtain
\begin{equation*}
\tilde{H}
=
p^{-} \dot{q}
+ p \dot{q}^{-}
-H
=
p^{-} \dot{q}
+ p \dot{q}^{-}
-  pp^{-} - qq^{-}
\end{equation*}

Delay equations~(\ref{HamEq32}) admit  symmetry operators
\begin{multline}
X_{1}=\sin t {\frac{\partial}{\partial q}} + \cos t {\frac{\partial}{\partial p}},
\qquad
X_{2}=\cos t {\frac{\partial}{\partial q}}-\sin t {\frac{\partial}{\partial p}},
\\
X_{3}={\frac{\partial}{\partial t}},
\qquad
X_{4}=q{\frac{\partial}{\partial q}}+p{\frac{\partial}{\partial p}},
\qquad
X_{5}=p{\frac{\partial}{\partial q}}-q{\frac{\partial}{\partial p}}.
\end{multline}
We note that symmetry  $X_{5}$ has no corresponding Lie point symmetry in the Lagrangian approach.

For symmetry $X_{1}$, the Hamiltonian is divergence invariant
\begin{equation*}
X_1 ( \tilde{H} ) + \tilde{H}   D ( \xi_1 )
=D ( \cos t^{-} q   +   \cos t \  q^{-}    ).
\end{equation*}
The symmetry gives the differential-difference relation
\begin{equation*}
D (  \sin t \  ( p^+ + p^- )   -   \cos t^{-} q -   \cos t  \  q^{-} )
= ( S_+  - 1 )  (  \cos t^{-}  \dot{q}    -   \sin t^{-} q ).
\end{equation*}
Since the right-hand side of the differential-difference relation can be rewritten as a total derivative
\begin{equation*}
( S_+  - 1 )  (  \cos t^{-}  \dot{q}    -   \sin t^{-} q )
= D (   \cos t \  {q^+}   - \cos t^{-}  {q} ),
\end{equation*}
we obtain the differential integral
\begin{equation}
I_1
=
 \sin t  \    ({p}^{+}+{p}^{-})
-    \cos t  \   (q^{+}+q^{-}).
\end{equation}


Symmetry  $X_{2}$ also satisfies divergence invariance of the Hamiltonian
\begin{equation*}
X_2 ( \tilde{H} ) + \tilde{H}   D ( \xi_1 )
= D(  -   \sin t^{-}  q    -    \sin t  \    q^{-}  ).
\end{equation*}
We obtain the  differential-difference relation
\begin{equation*}
D (  \cos t \  ( p^+ + p^- )   +    \sin  t^{-} q  +    \sin  t  \  q^{-} )
= ( S_+  - 1 )  (   - \sin t^{-}  \dot{q}    -   \cos  t^{-} q ).
\end{equation*}
Rewriting  the right-hand side of the  differential-difference relation as the total derivative
\begin{equation*}
 ( S_+  - 1 )  (   - \sin t^{-}  \dot{q}    -   \cos  t^{-} q )
= D ( - \sin  t  \ q^+  +  \sin  t^- q  ),
\end{equation*}
we obtain the  differential integral
\begin{equation}
I_2 =   \cos t   \     ({p}^{+}+{p}^{-})    +    \sin t    \
(q^{+}+q^{-}).
\end{equation}


Thought symmetry $X_{3}$  is a variational symmetry of the Hamiltonian,
i.e.,
\begin{equation*}
X _3 ( \tilde{H} ) + \tilde{H}   D ( \xi _3 ) =  0,
\end{equation*}
It needs other equations than the canonical Hamiltonian equation~(\ref{HamEq32}),(\ref{delays}) for deriving the differential-difference relation.


For symmetry  $X_{4}$, we find
\begin{equation*}
X _4 ( \tilde{H} ) + \tilde{H}   D ( \xi _4  )
= 2 (   p^-  \dot{q} +   p  \dot{q} ^-   - p p^- - q q^- ).
\end{equation*}
The symmetry is neither variational symmetry nor divergence symmetry of the Hamiltonian. Thus, there is no differential-difference relation.


Symmetry  $X_{5}$ is a divergence symmetry of the Hamiltonian
\begin{equation*}
X _5 ( \tilde{H} ) + \tilde{H}   D ( \xi _5 )
=  D (   p p^-  - q q^-  ).
\end{equation*}
It provides the differential-difference relation
\begin{equation*}
D (   p  p^+     +  q q^-   )
= ( S_+  - 1 )  (      p   \dot{p} ^- - q^- \dot{q} -  q  p ^-   +  q^- p   ).
\end{equation*}
Converting this relation to a differential or difference first integral is not possible.

\medskip

The differential first integrals $I_1$ and $I_2$ can be used to find solutions to the delay Hamiltonian equations~(\ref{HamEq32}). In fact, setting the values of these first integrals as
\begin{equation*}
I_1 = A,
\qquad
I_2 = B,
\end{equation*}
where $A$ and $B$ are constant, we obtain the following solution
\begin{subequations}   \label{sol23}
\begin{gather}
q^{+}+q^{-}= - A \cos t  + B \sin t, \\
{p}^{+}+{p}^{-}=  A\sin t + B  \cos t.
\end{gather}
\end{subequations}

Using this representation, one can construct a solution to a Cauchy problem with the initial values
\begin{equation*}
q(t) = \varphi(t),
\qquad p(t) = \psi(t),
\qquad
t \in [-2\tau, 0],
\end{equation*}
where the functions $ \varphi(t) $ and $ \psi(t) $ are assumed to be continuous.
By rewriting relations (\ref{sol23}) in shifted form
\begin{subequations}   \label{recursion}
\begin{gather}
q(t) + q(t - 2\tau) = - A \cos(t - \tau) + B \sin(t - \tau), \\
p(t) + p(t - 2\tau) = A \sin(t - \tau) + B \cos(t - \tau),
\end{gather}
\end{subequations}
we find the constants $ A $ and $ B $ as
\begin{equation*}
A = - \cos \tau  \left(\varphi(0) + \varphi(-2\tau)\right) - \sin \tau  \left(\psi(0) + \psi(-2\tau)\right),
\end{equation*}
\begin{equation*}
B = - \sin \tau  \left(\varphi(0) + \varphi(-2\tau)\right) + \cos \tau  \left(\psi(0) + \psi(-2\tau)\right).
\end{equation*}

Applying relations~(\ref{recursion}), we derive the solution of the Cauchy problem for $ t \in [0, 2\tau] $
\begin{equation*}
q(t) = - \varphi(t - 2\tau) - A \cos(t - \tau) + B \sin(t - \tau),
\end{equation*}
\begin{equation*}
p(t) = - \psi(t - 2\tau) + A \sin(t - \tau) + B \cos(t - \tau).
\end{equation*}
Then, for $ t \in [2\tau, 4\tau] $
\begin{equation*}
q(t) = \varphi(t - 4\tau) + A \cos(t - 3\tau) - B \sin(t - 3\tau) - A \cos(t - \tau) + B \sin(t - \tau),
\end{equation*}
\begin{equation*}
p(t) = \psi(t - 4\tau) - A \sin(t - 3\tau) - B \cos(t - 3\tau) + A \sin(t - \tau) + B \cos(t - \tau).
\end{equation*}
This process can be continued recursively. Using these relations, one can find the solution $ (q(t), p(t)) $ for $ t \in [0, \infty) $. Unlike the standard method of solving DODEs, namely the method of steps~\cite{bk:Elsgolts1955}, which requires integration, the recursive procedure outlined above does not involve any integration.

\section{Concluding remarks}

\label{section_Conclusion}

This paper provides a review of research on the application of Lie group transformations to delay ordinary differential equations.

First, DODEs of first- and second-order DODEs with a single delay were classified into conjugacy classes, under arbitrary Lie point transformations. Here, we examine the invariance of a system of two equations: the DODE itself and the equation specifying a delay. The method for obtaining particular solutions of DODEs by symmetry reduction was presented.

Then, the Lagrangian approach for variational delay ordinary differential equations is presented for first-order delay Lagrangians. The variational approach to constructing DODEs necessarily leads to equations with two delays. Noether-type operator identity relates the invariance of a delay functional with the appropriate variational equations and their conserved quantities. This identity is used to state a Noether-type theorem that gives first integrals of variational second-order DODEs with symmetries.

Finally, the Hamiltonian approach to delay differential equations is described. The delay analog of the Legendre transformation, which relates the Lagrangian and Hamiltonian approaches, is discussed.
This transformation possesses several coefficients that should be appropriately related to the form of the Lagrangian. The transition from a Lagrangian function to the corresponding Hamiltonian function has been demonstrated for Lagrangians that are quadratic in the derivatives. The delay analog of the Hamiltonian operator identity,  which relates the invariance of a delay functional with the appropriate variational equations and their conserved quantities, is established. The Noether-type theorem was derived from this identity. In particular, it holds for delay canonical Hamiltonian equations with constant delay, with symmetries acting in the space of the dependent variables.

For both the Lagrangian and Hamiltonian approaches, first integrals can be used to express solutions. In cases of sufficiently many first integrals, the solutions of the DODEs can be presented recursively.

In the paper, we did not explain how the invariance of variational equations relates to the invariance of delay functionals. This topic was considered for the Lagrangian approach in \cite{DKM2023} and for the Hamiltonian approach in \cite{DKM2025}.

In conclusion, we formulate a range of problems that still await solution.

\begin{enumerate}

\item

S.~Lie formulated an algorithm for the complete integration of second-order ODEs with two or more symmetries \cite{Lie_1091}. It would be tempting to develop a similar method for second-order DODEs with the required number of symmetries.

\item

The so-called direct method, which allows finding conservation laws (first integrals) of differential equations without using symmetries, has been significantly developed in works \cite{bk:AncoBluman1997, Anco_1998, bk:BlumanAnco2002}. It would be tempting to generalize this method to differential equations with retarded arguments.

\item

The adjoint equation method, based on the Lagrange operator identity, allows one to find conservation laws (first integrals) of differential equations, regardless of whether a variational formulation exists for these equations. The works of G.~Bluman and S.~Anco \cite{Anco_2002a, Anco_2002b, bk:BlumanAnco2002} made significant contributions to this method. This method has been successfully generalized to ordinary difference equations \cite{DKKW2015}. Hopefully, it can be generalized to DODEs.

\item

The variational approach to constructing equations with delay inevitably leads to equations with two (or more) delay parameters. However, a group classification of differential equations with two delay parameters has not yet been completed. This task is difficult even for DODEs with constant delay parameters, as many variables appear in the equation's description. This problem becomes even more complicated if the classification is carried out under the assumption that the delay parameter may depend on the solution.

\item

In practical problems involving mathematical models with a delay argument (e.g., chemistry, biology, etc.), equations with a single constant delay parameter are most often used. However, these models are not derived from the assumption of any conserved quantities or integrals. Meanwhile, modeling processes based on delay conservation laws may yield equations with two delay parameters and the possibility of variational formulations.

\end{enumerate}

\section*{Dedication}

The paper is dedicated to George Bluman, who made significant contributions to the application of Lie groups to differential equations.
The authors are grateful to him for  introducing  the use of the Lagrange operator identity to find conservation laws (first integrals) for non-variational equations that admit symmetries.


\label{lastpage}
\end{document}